\documentclass[fleqn,12pt,twoside]{article}
\usepackage[headings]{espcrc1}
\usepackage{floatflt}
\readRCS
$Id: espcrc1.tex,v 1.2 2004/02/24 11:22:11 spepping Exp $
\usepackage{color}
\usepackage{graphicx}
\usepackage[figuresright]{rotating}
\title{Non-identical particle correlations at 62 and 200 GeV at STAR}

\author{Petr Chaloupka  \address[ujf]{Nuclear Physics Institute, Academy of Sciences
        of the Czech Republic, 250 68 Rez near Prague, Czech Republic}
        (for the STAR\thanks{For the full list of STAR authors and acknowledgments,
          see appendix 'Collaborations' of this volume.}
         collaboration) 
      }

\runauthor{P. Chaloupka}

\begin{document}

\maketitle

\begin{abstract}
We report on STAR analyses of 
\mbox{$p\!-\!\Lambda$}, \mbox{$\bar{p}\!-\!\bar{\Lambda}$},
\mbox{$p\!-\!\bar{\Lambda}$}, \mbox{$\bar{p}\!-\!\Lambda$},
$p\!-\!\bar{p}$ and $\pi\!-\!\Xi$ correlations
in Au+Au collisions at $\sqrt{s_{NN}}=62$
and $200$~GeV.
Measured source sizes in 
\mbox{$p(\bar{p})\!-\!\Lambda(\bar{\Lambda})$}
and $p\!-\!\bar{p}$ are shown to be in qualitative agreement with flow expectations.
Interaction potential between
\mbox{$p\!-\!\bar{\Lambda}$}, \mbox{$\bar{p}\!-\!\Lambda$}
was investigated by measuring scattering length, 
thus showing that
correlation analyses in heavy-ion collisions can be used to study
strong interaction potential between hadrons.
We present also analyses of $\pi\!-\!\Xi$ correlations addressing
independently on previous measurements the issue of $\Xi$ flow in heavy-ion collisions.
\end{abstract}

\section{Introduction}
\vspace{-0.2cm}
Measurements of momentum correlation of particles at small
relative velocities are used to study space-time characteristics
of the heavy-ion collisions~\cite{Lisa_Pratt_review}.
Both identical and non-identical particle correlations
are sensitive to the space-time extent
of the particle-emitting source.
However, as suggested in~\cite{Lednicky}, non-identical particle correlation
measurements provide additional information about  relative 
space-time emission asymmetry among the two particles. 

Current data
from Au+Au collisions~\cite{QM2005_flow_200GeV}
suggest that the hot and
dense system created in the heavy-ions collision
builds up substantial collectivity
leading to a rapid transverse 
expansion.
Flow induces
a strong correlation between particles' velocities and emission
points
leading to an effective decrease of measured HBT radii and 
different average emission points for particle species with non-equal
masses~\cite{blastwave}.
Therefore non-identical correlations can be used as an independent 
cross-check  of flow measurements in heavy-ion collisions.

Furthermore since the correlations between non-identical hadrons 
arise from
their final state strong and, for charged particles,
Coulomb interaction, these measurements can be used
to study \cite{Lisa_Pratt_review,Lednicky} 
strong interaction potentials between particles which could otherwise
be hardly accessible by other means.

\section{Analyses methods, Experimental data }
\vspace{-0.2cm}
The correlation between non-identical particles depends on
$\vec{k^*}=\vec{k_1^*}=-\vec{k_2^*}$, which is
the first particles' momenta in the pair's rest frame,
{\it i.e.} half of the momentum difference between the particles.
A small value of $|{\vec{k^*}}|$ then 
means that the particles move
with a small relative velocity.
The correlations are studied,
as in previous STAR measurements~\cite{Kisiel_QM2004}
by constructing correlation functions
$C(\vec{k^*})=A(\vec{k^*})/B(\vec{k^*})$,
as a ratio of two-particle distributions;
$A(\vec{k^*})$ - obtained from single event,
and one where particles come from different, or
  "mixed" events - $B(\vec{k^*})$.

STAR main detector, the Time Projection Chamber (TPC), detects and reconstructs
charged particles emerging from primary and secondary vertices.
Pions, kaons and protons are identified via their specific energy
loss (dE/dx). 
This selection limits 
transverse momentum acceptance of  pions
to $0.08 < p_t < 0.6$~GeV/$c$, 
and of protons to $0.4 < p_t < 1.1$~GeV/$c$.
Lambdas (and anti-lambdas)
 as well as charged $\Xi$-hyperons
are topologically reconstructed using decay chain
$\Xi\rightarrow\Lambda+\pi$, and 
$\Lambda\rightarrow\pi+p$.
Only events with longitudinal primary vertex position within $25$~cm
of the TPC center, and only particles in the rapidity
window $|y| < 0.5$ are selected.
This subsequently limits the $p_t$ range of our lambda-sample to 
\mbox{$0.3 < p_t < 2.0$ GeV/$c$} and \mbox{$\Xi$-sample to $0.7 < p_t < 3.0$ GeV/$c$}.

\section{Results}
\begin{figure}[t]
\begin{minipage}[h]{0.49\textwidth}
\centering
\includegraphics[width=0.95\textwidth,clip]{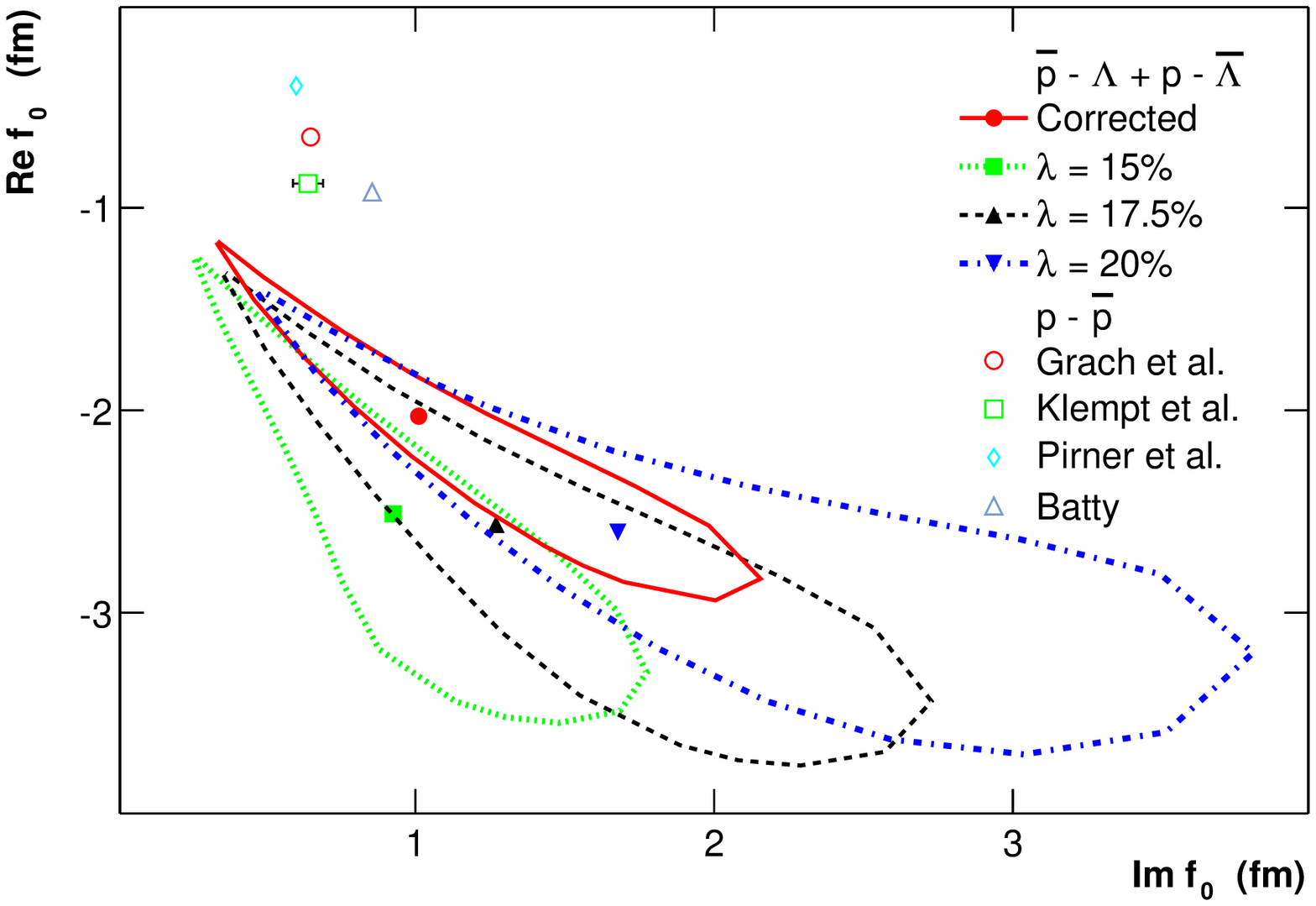}
\vspace{-0.4cm}
\caption{Spin-averaged s-wave scattering length (preliminary) for
 \mbox{$p\!-\!\bar{\Lambda}+\bar{p}\!-\!\Lambda$} with one standard 
 deviation contours compared to previous \mbox{$p\!-\!\bar{p}$} measurements.}
\label{fig:chi2}
\end{minipage}
\hfill
\begin{minipage}[h]{0.49\textwidth}
\centering
\includegraphics[width=\textwidth,clip]{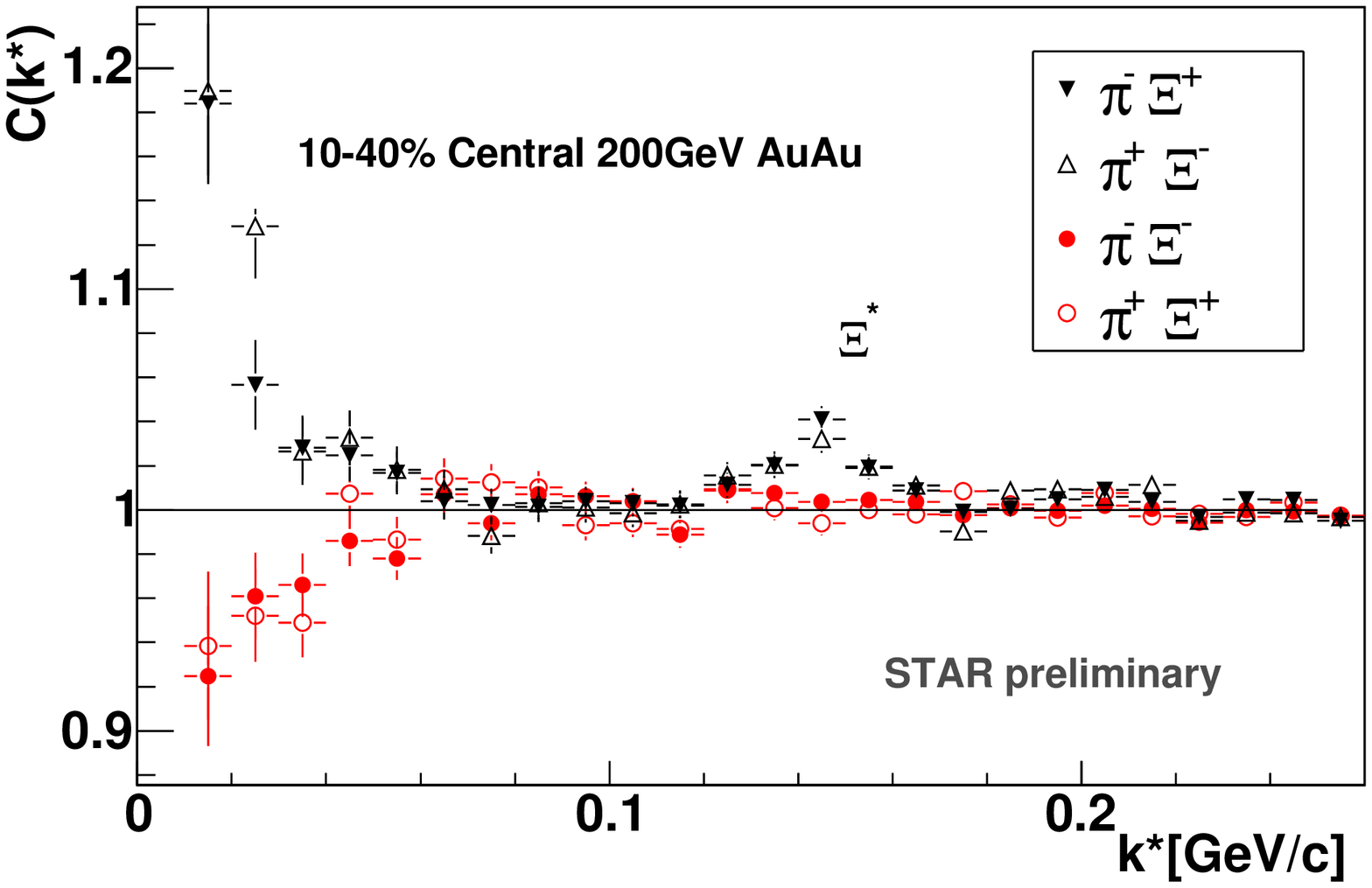}
\vspace{-0.8cm}
\caption{\mbox{$\pi\!-\!\Xi$} correlation function for 10\% most
  central Au+Au collisions at 
\mbox{$\sqrt{s_{NN}}$=200}~GeV.
  Peak at \mbox{$k^*=0.15$ GeV/$c$} corresponds to $\Xi^*(1530)$ resonance.}
\label{fig:piXi}
\end{minipage}\hfill
\end{figure}
STAR has performed correlation measurements in
\mbox{$p(\bar{p})-\Lambda(\bar{\Lambda})$} and $p\!-\!\bar{p}$ systems.
We refer reader to Figure~3 in~\cite{QM2005_idHBT} in these proceedings,
where we present the preliminary results on $m_T$ dependence of
extracted
one-dimensional Gaussian
 source radii for different systems including results for
\mbox{$p\!-\!\Lambda$}, \mbox{$\bar{p}\!-\!\bar{\Lambda}$}, 
\mbox{$p\!-\!\bar{\Lambda}$}, \mbox{$\bar{p}\!-\!\Lambda$}, 
and $p\!-\!\bar{p}$ from
10\% most central Au+Au collisions at \mbox{$\sqrt{s_{{NN}}}=200~GeV$}.
\begin{figure}[t]
\begin{minipage}[h]{0.49\textwidth}
\centering
\includegraphics[width=\textwidth,clip]{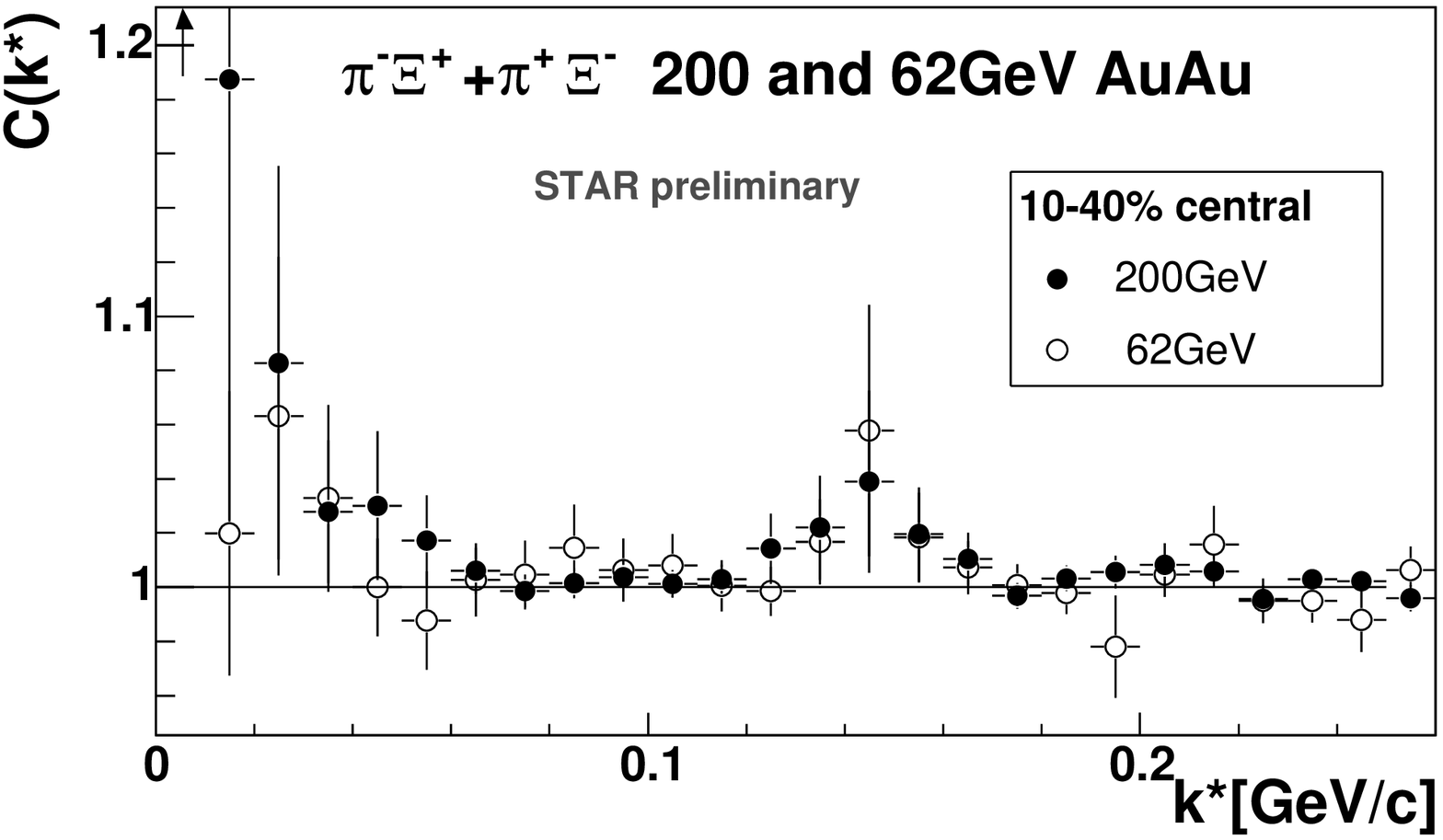}
\includegraphics[width=\textwidth,clip]{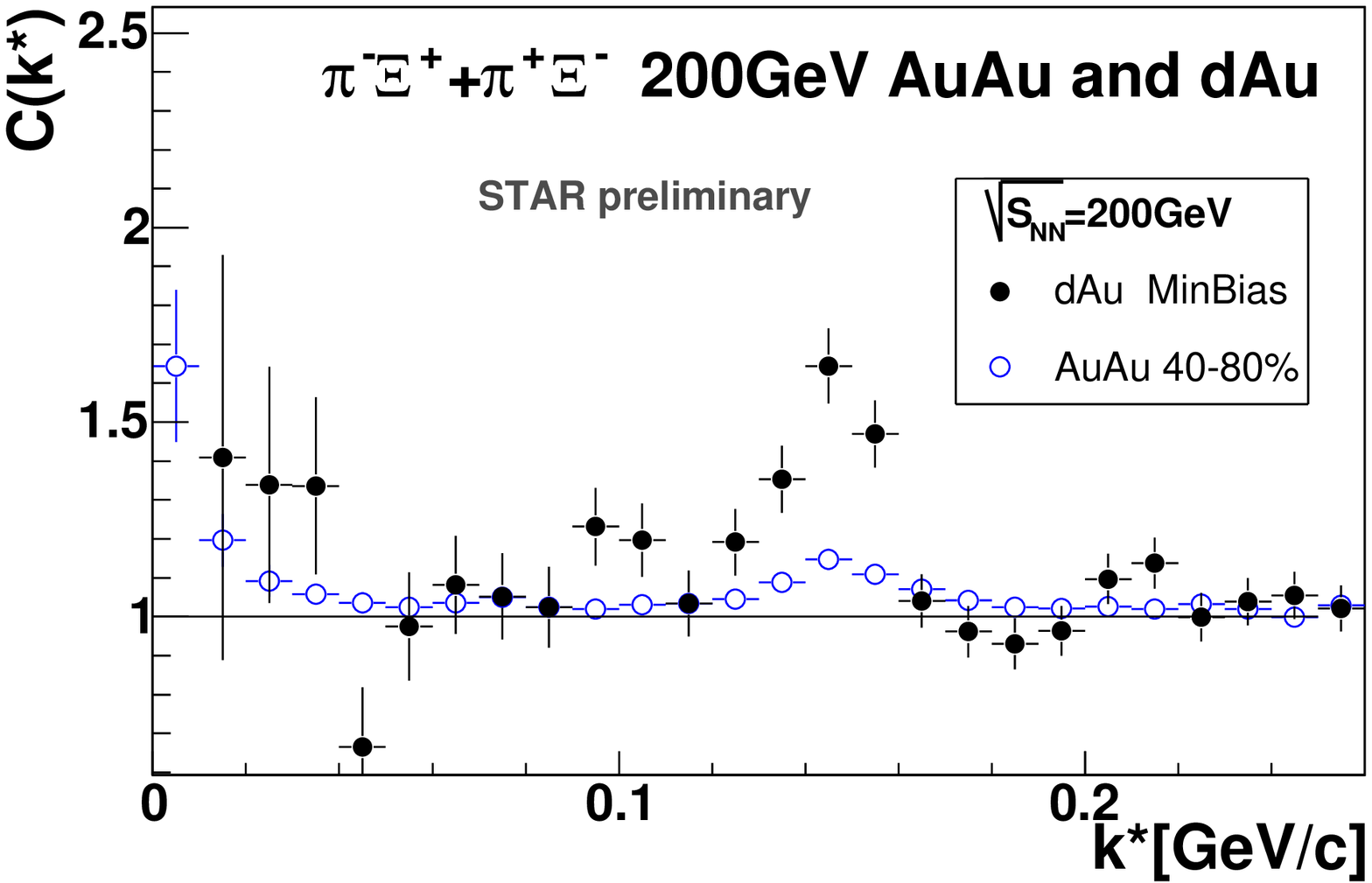}
\vspace{-0.45cm}
\caption{Comparison of combined unlike-sign $\pi\!-\!\Xi$
  $C(k^*)$:
  Top - for two different energies in Au+Au collisions;
  Bottom - for Au+Au and d+Au collisions at 200~GeV.
}
\label{fig:piXi_comp}
\end{minipage}
\hfill
\begin{minipage}[h]{0.49\textwidth}
\centering
\includegraphics[width=\textwidth,clip]{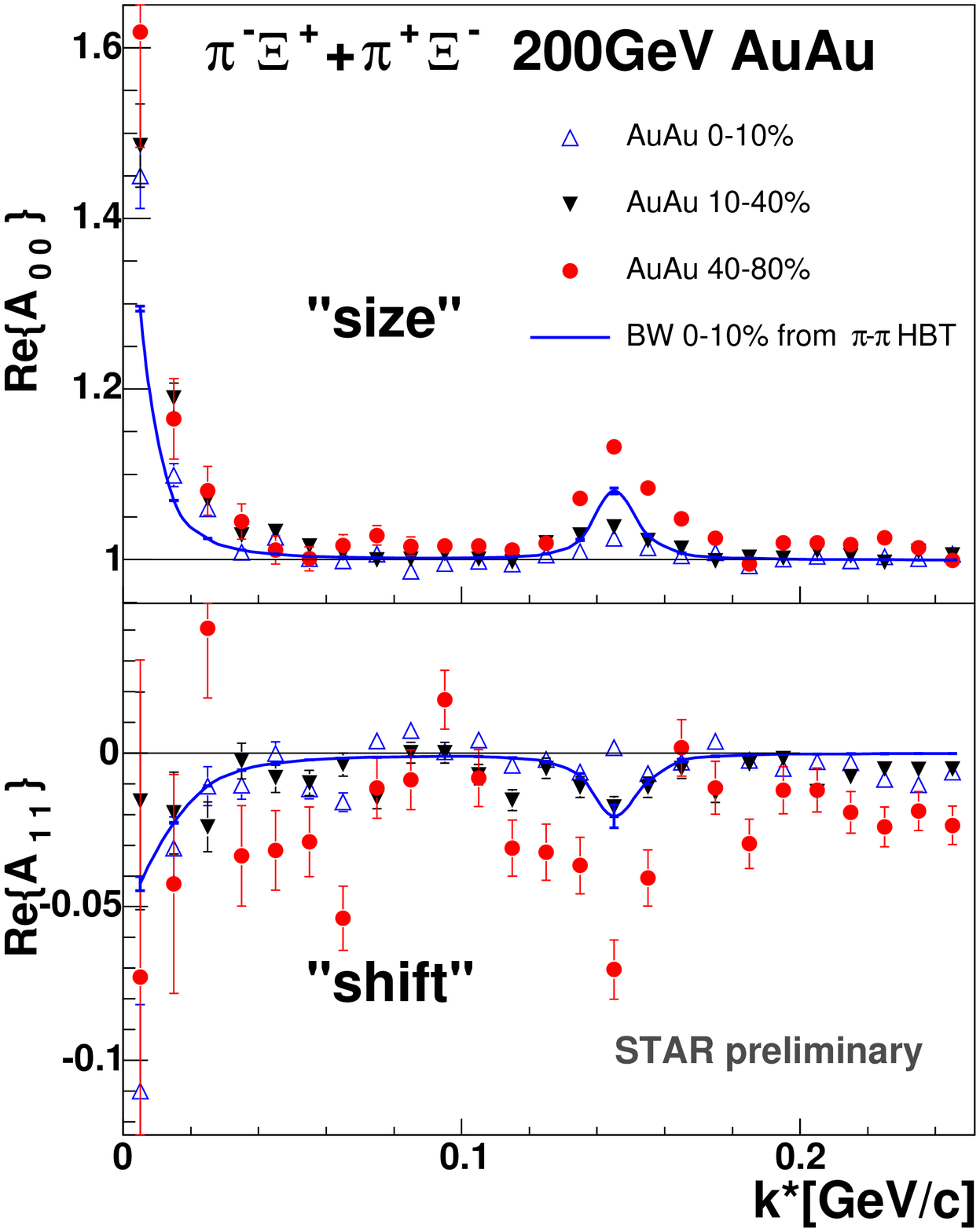}
\vspace{-1.1cm}
\caption{Combined unlike-sign $\pi\!-\!\Xi$ pairs:
  centrality dependence of
  spherical projections of $C(\vec{k^*})$ in 200~GeV Au+Au.
  Solid line is a theoretical prediction for the most central data
  assuming $\Xi$ flow.}
\label{fig:SH}
\end{minipage}\hfill
\end{figure}
The \mbox{$p(\bar{p})\!-\!\Lambda(\bar{\Lambda})$}
radii were extracted by fitting an analytical form of $C(k^*)$
using Lednick\'y \& Lyuboshitz
final state interaction (FSI) model~\cite{lednicky_lub_model}.
While the interaction potential for 
$p\!-\!\Lambda$ and $\bar{p}\!-\!\bar{\Lambda}$ is known, allowing   
us to extract the source size, the correlation function for 
$\bar{p}\!-\!\Lambda$ and $p\!-\!\bar{\Lambda}$ was measured for the first time,
and the interaction is unknown. 
In the case of \mbox{$\bar{p}\!-\!\Lambda+p\!-\!\bar{\Lambda}$}
the correlation function
was fitted assuming the same 
functional form of the interaction
as in
$p\!-\!\Lambda$, $\bar{p}\!-\!\bar{\Lambda}$, treating the potential
parameters (scattering lengths) as free parameters.
The extracted spin-averaged scattering lengths
are presented in Figure~\ref{fig:chi2} for different values of 
used pair purity corrections~$\lambda$. 
The best results are obtained when $k^*$ dependent
purity correction is applied (the ``Corrected'' curve).
In order to take into account annihilation channels the scattering length 
has to have a significant imaginary part.
As can be seen, the extracted radii for
  $\bar{p}\!-\!\Lambda$, $p\!-\!\bar{\Lambda}$
differ from those extracted from $p\!-\!\Lambda$ and $\bar{p}\!-\!\bar{\Lambda}$.
This difference can possibly be explained by a correlated feed-down,
which has not been accounted for, and is now under study.

STAR has also measured $p\!-\!\bar{p}$ correlations at \mbox{$\sqrt{s_{NN}}=62.4$~GeV}  
and at different centralities with comparable result~\cite{QM2005_Poster_p-p_Gos}. 
In $p\!-\!\bar{p}$ measurements correlation functions were fitted
numerically \cite{corrfit} 
using the same FSI model \cite{lednicky_lub_model} as 
in \mbox{$p(\bar{p})\!-\!\Lambda(\bar{\Lambda})$} case.

The relatively smaller source radii obtained from 
\mbox{$p(\bar{p})\!-\!\Lambda(\bar{\Lambda})$} and $p\!-\!\bar{p}$, as function of $m_T$,
when compared to lighter particles agree with effects expected from flow~\cite{blastwave}.

In Figure~\ref{fig:piXi} we present the preliminary results on
$C(k^*)$ for all four combinations of \mbox{$\pi\!-\!\Xi$}
pairs from the 10\% most central $200$~GeV Au+Au collisions.
In the low $k^*$ region \mbox{($k^* < 0.05$ GeV/$c$)} the correlation
function is, for all charge combinations, dominated by the Coulomb interaction. 
Final state strong interaction is manifested in $C(k^*)$  of unlike-sign 
pairs as a peak at \mbox{$k^*=0.15$~GeV/$c$}, corresponding
to $\Xi^*(1530)$ resonance.

In Figure~\ref{fig:piXi_comp} comparison of purity corrected $C(k^*)$ for
combined unlike-sign \mbox{$\pi\!-\!\Xi$} pairs from Au+Au
collisions at $\sqrt{s_{NN}}=62.4$~GeV and $200$~GeV is shown
together with comparison of Au+Au and d+Au collision systems at \mbox{$\sqrt{s_{NN}}=200$~GeV}.
All correlation functions exhibit the same general features
in all presented systems and energies.

Recent data suggest that $\Xi$ develops substantial elliptic flow
during heavy-ion collisions~\cite{QM2005_flow_200GeV}.
Independent test of this hypothesis can be pursued 
via decomposition of $C(\vec{k^*})$ into spherical harmonics~\cite{Chajecki_Lisa_SH}.
In Figure~\ref{fig:SH} we present centrality dependence of 
$A_{00}(k^*)$ and $A_{11}(k^*)$ components for \mbox{$\sqrt{s_{NN}}=200$~GeV} Au+Au collisions.
In the same plot is shown a prediction for the most central data based on 
S.~Pratt's model~\cite{Pratt_model}, where blastwave model was used to provide emission 
space-time coordinates taking into account the influence of flow.
Blastwave parameters describing well Au+Au data were used for both particles,
assuming significant $\Xi$ flow. 
While $A_{00}(k^*)$ is angularly averaged $C(\vec{k^*})$,
$A_{11}(k^*)$ is sensitive to emission asymmetry in the system,
vanishing in a system where both particles are emitted on average at the same
space-time point.

From Figure~\ref{fig:SH}~and~\ref{fig:piXi_comp} we observe that $\pi\!-\!\Xi$ 
correlation function shows strong centrality  and system dependence,
while it doesn't seem to be significantly sensitive to the collision energy.
Moreover in the region of $\Xi^*(1530)$ the
$C(k^*)$ shows much stronger sensitivity to the source size than in the
Coulomb region.
The $A_{11}(k^*)$ differs significantly from zero in both,
Coulomb and $\Xi^*(1530)$, regions qualitatively following 
theoretical prediction,
thus implying that pions and $\Xi$s are not emitted
from the same average space-time point,
 and supporting evidence of $\Xi$ experiencing 
transverse radial flow.

\section{Conclusions}
We have presented preliminary results on 
\mbox{$p(\bar{p})\!-\!\Lambda(\bar{\Lambda})$},
$p\!-\!\bar{p}$, and $\pi\!-\!\Xi$ correlation measurements 
from STAR experiment at RHIC. 
In \mbox{$p(\bar{p})\!-\!\Lambda(\bar{\Lambda})$} and $p\!-\!\bar{p}$
extracted emission radii are consistent 
with transversally expanding particle source.
In case of \mbox{$p\!-\!\bar{\Lambda}$}, \mbox{$\bar{p}\!-\!\Lambda$},
not only size, but also scattering length was extracted from the fit for the first time,
showing that non-identical correlations can be used to study
hadron interactions directly.
We have also presented preliminary results on $\pi\!-\!\Xi$ correlations
at two energies in Au+Au and d+Au collisions showing sensitivity of the $C(\vec{k^*})$
to the size of the system - mainly in the region of $\Xi^*$ resonance peak.
We have used novel technique of spherical decomposition to 
present independent evidence of $\Xi$ flow in heavy-ion collisions.

\end{document}